\documentclass[journal=jctc,manuscript=article]{achemso}
\usepackage[utf8]{inputenc}
\usepackage{amsmath,color}
\usepackage{amsfonts}
\usepackage{mathrsfs}
\usepackage{mathtools}
\usepackage{graphicx}
\graphicspath{ {./images/} }
\numberwithin{equation}{section}

\DeclarePairedDelimiter\bra{\langle}{\rvert}
\DeclarePairedDelimiter\ket{\lvert}{\rangle}
\DeclarePairedDelimiterX\braket[2]{\langle}{\rangle}{#1 \delimsize\vert #2}
\DeclarePairedDelimiterX\Braket[3]{\langle}{\rangle}{#1 \delimsize\vert #2 \delimsize\vert #3}

\SectionNumbersOn

\newcommand{\pd}[2]{\frac{\partial #1}{\partial #2}}

\title{Electron Transfer and Spin Orbit Coupling: How Strong are Berry Force Effects In and Out of Equilibrium In the Presence of Nuclear Friction?}

\author{Suraj Chandran}
\author{Yanze Wu}
\author{Hung-Hsuan Teh}
\author{David H. Waldeck}
\author{Joseph E. Subotnik}

\email{surajc@sas.upenn.edu, wuyanze@sas.upenn.edu, teh@sas.upenn.edu, dave@pitt.edu, subotnik@sas.upenn.edu}

\begin{document}
\maketitle

\begin{abstract}
    We investigate a spin-boson inspired model of electron transfer, where the diabatic coupling is given by a position-dependent phase, $e^{iWx}$.  We consider both  equilibrium and nonequilibrium initial conditions. We show that, for this model, all {\em equilibrium}  results are completely invariant to the sign of $W$ (to infinite order).  However,  the {\em nonequilibrium} results do depend on the sign of $W$, suggesting that  photo-induced electron transfer dynamics are meaningfully affected by Berry forces even in the presence of nuclear friction; furthermore, whenever there is spin-orbit coupling, electronic spin polarization can emerge (at least for some time).  
\end{abstract}


\section{Introduction}\label{sec:intro}

Quantum mechanical models of electron transfer have proven extremely successful in describing key features of electronic transitions in systems such as biomolecules and solar cells. One of the simplest models that has been effective in such applications is the spin-boson model, where a two-state system (representing two electronic states) is coupled to a thermal bath (represented by a collection of harmonic oscillator modes)\cite{leggett:1987:rmp}. In the most basic formulation, one assumes that the two states are coupled together via a constant coupling ($V$), an assumption known as the Condon approximation. If the set of modes in the thermal bath is indexed by $\alpha$, the Hamiltonian of such a system is of the form
\begin{eqnarray}\label{condon}
\label{eqveryfirst}
    \hat H &=& \hat H_0 + 
    \hat V \\
    \hat V & = & V\ket1\bra2+V^*\ket2\bra1\\
    \label{joeH0}
    \hat H_0 &= &
    \left(E_1+\sum_\alpha c^{(1)}_{\alpha}\hat x_\alpha\right)\ket1\bra1+\left(E_2+\sum_\alpha c^{(2)}_{\alpha}\hat x_\alpha\right)\ket2\bra2 + \hat H_B \\
\hat H_B&=&\sum_\alpha \frac{\hat p_\alpha^2}{2m_\alpha}+\frac12m_\alpha\omega_\alpha^2\hat x_\alpha^2
\end{eqnarray}

Within the context of chemical physics, the spin-boson problem has proved itself useful as a model because (i) the Hamiltonian is simple enough such that one can solve for the dynamics at many different levels of theory, and (ii) there are so many time scales of interest (temperature, driving force, diabatic coupling, reorganization energy, and the nuclear relaxation time) that one can explore many different dynamical regimes\cite{jortner:1999:advchemphys}. For instance, in the nonadiabatic limit, where V is small, the electronic dynamics become slow enough that the nuclear dynamics cannot be sufficiently described using the Born-Oppenheimer approximation and the usual approach is to apply the Fermi Golden Rule. Thereafter, if one makes the high temperature approximation, one recovers what is often called Marcus theory\cite{nitzanbook}. In the adiabatic limit of the spin-boson problem, the system evolves as if on a single electronic surface and one can invoke standard classical transition state theory (TST) to approximate the rate of reaction\cite{marcus:1956}. Yet another limit of the spin-boson model problem is the solvent-controlled regime where nuclear motion can slow down electronic transitions by trapping transitions with strong friction\cite{zusman:1980:original,hynes:1986:zusman,straub:1987:berne_nonadiabatic}.

Extensions of the spin-boson model Hamiltonian have been proposed and analyzed as well.  For instance, several researchers have explored how Marcus theory is altered in the presence of position-dependent diabatic couplings (breaking the so-called ``Condon'' approximation). Perhaps the most famous example of such a model is due to Stuchebrukhov\cite{stuchebrukhov:1997:noncondon}, who studied systems where all nuclear vibrational modes could be disentangled into two groups: modes that displace the diabat and modes that modulate the diabatic coupling. In such a regime,  a simple rate expression can be derived. These expressions were effectively generalized by Jang and Newton\cite{jang:2005:noncondon}, who calculated the consequences of non-Condon effects for a variety of different circumstances. 

Below, our focus will be slightly different from the papers mentioned above insofar as our aim
will be to model the effect of spin-orbit coupling on nonadiabatic electron transfer, a subject that has recently been reviewed by Varganov {\em et al}\cite{varganov:2016:ijqc} and Marian {\em et al} \cite{marian:2018:chemrev:spinvibronic}. In order to treat systems with spins as rigorously as possible, we will not force the diabatic coupling to be real-valued. Instead, we will explore the implications of the fact that the $\hat{\textbf L} \cdot \hat{\textbf S}$ spin-orbit coupling operator\cite{marian:2001:revcompchem,marian:2012:wires_isc} is complex-valued. The result is that a spin-conserving process between two electronic states (say,with spin up) evolves according to a spin-dependent Hamiltonian $\hat H$;  while the dynamics of the corresponding process (say, with  spin down) evolve according to a spin-dependent Hamiltonian $\hat H^*$. A brief justification of this model is provided in Appendix \ref{LS_app}.

Although not widely appreciated in the chemistry community, when nuclei are propagated according to a complex-valued Hamiltonian, a novel so-called Berry force\cite{berry:1993:royal:half_classical,subotnik:2019:jcp_berry_qcle,hedegard:2010:nano_berry} emerges from the fact that no consistent phase can be chosen for the electronic adiabats.  Historically, the Berry force was derived by Robbins and Berry\cite{berry:1993:royal:half_classical} in the adiabatic limit of nuclear motion. Interestingly, recent dynamical scattering simulations of systems with Berry force have demonstrated that large Berry forces can also appear in systems with a few degrees of freedom if wavepackets reach geometries near conical intersections.\cite{wu:2020:neartotal} However, to our knowledge,  the implications of Berry force have not been analyzed in the context of Marcus theory, and there have been few if any calculations investigating how Berry force effects are modulated or controlled in the presence of nuclear friction. The present article makes a first step in this direction by using the spin-boson model to ask the question: how important will Berry force effects be for electron transfer in the condensed phase?

Because exact quantum dynamics is difficult to propagate in the condensed phase, a simplifying assumption will be necessary;  we will make the approximation that the diabatic coupling is small so that one can use first order perturbation theory and employ the Fermi Golden Rule (FGR) ansatz to calculate equilibrium rates:
\begin{eqnarray}\label{FGR}
k = \frac{2\pi}\hbar\sum_{\textbf v'}  \lvert\Braket{2,\textbf v'}{\hat V}{1, \textbf v}\rvert^2\delta\left(E_{2,\textbf v'} - E_{1,\textbf v}\right)
\end{eqnarray}
In Eq. \ref{FGR}, $\hat V$ can be a function of the nuclear bath coordinates $\left\{ x_{\alpha} \right\}$.   This expression gives the rate of population decay from an initial vibronic state $\ket{1,\textbf v}$ to a set of final states $\{\ket{2,\textbf v'}\}$. Here, $1$ and $2$ index electronic states and $\textbf v$ and   $\textbf v'$ index vibrational states. Now, if we assume that $(i)$ the unperturbed Hamiltonian ($\hat H_0$ in Eq. \ref{joeH0}) is real and $(ii)$ the  initial state $\ket{1, \textbf v}$ is a stationary state of the unperturbed Hamiltonian $\hat H_0$, it follows then that $\Braket{2, \textbf v'}{\hat V}{1,\textbf v}^* = \Braket{2, \textbf v'}{\hat V^*}{1,\textbf v},$ i.e. $\left|\Braket{2, \textbf v'}{\hat V}{1,\textbf v}\right|^2 = \left|\Braket{2, \textbf v'}{\hat V^*}{1,\textbf v}\right|^2.$ In other words,  to first order in $\hat V$ or $\hat V^*$, the rate of electronic relaxation for $\hat H$ must equal the rate of relaxation for $\hat H^*$.   Put bluntly, if the dynamics are initiated from quasi-equilibrated starting conditions, no Berry phase effect can arise to first order in perturbation theory. 

With this background in mind, it becomes clear that if Berry phase  effects are to arise, they must enter either from from higher orders in perturbation theory or from nonequilibrium starting conditions.  With regards to the former, we note that higher order effects can and do have important dynamical consequences. For instance, the famous Zusman result\cite{zusman:1980:original,hynes:1986:zusman,straub:1987:berne_nonadiabatic} is a higher-order effect beyond FGR demonstrating that, with enough friction, electron transfer rates can be dominated by solvent relaxation even for small diabatic couplings. Thus, in the future, we expect that an interesting research direction will be probing if/how Berry phase effects emerge at second or higher order in the perturbation.

For the present article, our focus will be on the latter possibility, i.e. the possibility that Berry force effects may arise from  {\em nonequilibrium} initial starting conditions. Such conditions are relevant to a great many electron transfer processes which occur before equilibration or a steady-state can be reached.  This scenario is true particularly in the study of photo-induced electron transfer, where the timescale of light-induced electronic excitation is much faster than that of nuclear vibrational relaxation\cite{barbara:1996:marcus}. Intuitively, we might expect  the phases of wavepackets take on greater significance for a complex-valued Hamiltonian, and thus coherences may play a role in amplifying Berry force effects. In order to understand the short-time behavior of relaxation in such systems, equilibrium methods are not sufficient. Fortunately, nonequilibrium systems can be physically modeled through a nonequilibrium formulation of Fermi's Golden Rule\cite{coalson:nitzan:1994:jcp:nonequilibrium}, a brief review of which will be provided in the following section.

At this point, all that remains is to introduce the spin-boson like complex-valued Hamiltonian that we will study to explore electron transfer with spin.  After investigating several different possibilities, we have settled on the following exponential spin-orbit coupling (ESOC) (with $\hat H_0$ as in Eq. \ref{joeH0}):
\begin{equation}\label{ESOC}
 \hat V = Ve^{i\sum_\alpha W_\alpha\hat x_\alpha} \ket1\bra2+
 V^*e^{-i\sum_\alpha W_\alpha\hat x_\alpha} \ket2\bra1\end{equation}
The interstate coupling in Eq.  \ref{ESOC} remains bounded for all displacement of the system (i.e. $\left|\hat V \right|$ does not diverge even when $\left|x_{\alpha}\right| \rightarrow \infty$)  making this ESOC model particularly amenable to theoretical investigation. More precisely, we will be able to apply FGR and NEFGR (non-equilibrium FGR) techniques to this Hamiltonian and gain basic insight into how Berry force does or does not manifest in molecular electron transfer rates. 
Note that a change in the sign of the $W_{\alpha}$ parameters corresponds to a change in the spin and this aspect will be discussed often. Again, see Appendix \ref{LS_app}. 

Lastly, as a motivation for the present research, it is worthwhile to put the present results in the context of the larger phenomenon of chiral induced spin selectivity (CISS)\cite{naaman:2015:arpc,naaman:2019:natrev}. Over the last 15 years, a host of experiments have shown that the electronic current running through a chiral molecule is often quite spin polarized\cite{naaman:2011:science:ciss_dna}. At the moment, there is no uniformly agreed upon explanation for this phenomena, given the very small magnitude of the spin-orbit coupling and Zeeman effects within simple organic molecules\cite{hermann:2020:jctc:ciss_est,mujica:2015:jcp_ciss,gersten_nitzan:2013:jcp_chiral}.  Over the last two years, our group\cite{wu:2020:jpca:spin} and the Fransson group\cite{fransson:2020prb:vibrational} have argued that the CISS effect must be due (at least in part) to nuclear-electronic interactions, i.e. the entanglement between  electronic transitions, spin-dependent Berry forces and nuclear motion. Indeed, the goal of Ref. \cite{wu:2020:neartotal} was to show that Berry forces can lead to strongly spin-dependent nuclear motion even for systems with small spin-orbit coupling. In this work, we will further show that such Berry force effects can also survive friction for some period of time (which is consistent with recent calculations of spin polarization emerging at a molecular junction under bias\cite{teh:wenjie:ef_dynamics_spin:2022} in the presence of {\em electronic} friction as caused by the production of electron-hole pairs). Although not indicative of any causation, this finding of robust spin effects is consistent with observations of the CISS effect at ambient temperatures for a wide range of molecular and solid state environments.
    
Before concluding this section, a word about language is in order. So far, we have very loosely spoken of ``Berry force effects'' as being those differences  that arise  when simulating  nuclear dynamics with Hamiltonian $\hat H$ versus  Hamiltonian ${\hat H^*}$.  This definition is not unique and not very precise.  First, not all differences between $\hat H$ and $\hat H^*$ dynamics can be attributed to Berry forces. Berry forces are pseudo-magnetic fields that arise in many-dimensional systems, whereas dynamical differences between $\hat H$ and $\hat H^*$ can arise even in one nuclear dimension; in fact, $\hat H$ and $\hat H^*$ can yield different electronic dynamics without any nuclear motion at all. Conversely, one can find  chemical circumstances where Berry force (or Berry phase) effects appear even with real-valued Hamiltonians ($\hat H = \hat H^*$) \cite{guo_yarkony:2016:pccp_review,guo_yarkony:2016:jacs,footnote:truhlar}.  For instance, theorists have long been interested in modeling the nuclear consequences of Berry phase around real-valued conical intersections.\cite{izmaylov:2015:fssh_ci_whysowell, izmaylov:2017:geometric_phase, fedorov:levine:discontinuous_basis, meek:levine:BO_conical, yarkony:review:conicalbook, yarkony:review2:conicalbook,yarokny:1996:rmp}
Berry forces are also known to emerge in the context of singlet-triplet crossings with real Hamiltonians\cite{bian:2021:perspective}. In short, it is clear that Berry force effects are not exactly equivalent to the differences between $\hat H$ and $\hat H^*$ dynamics.

Nevertheless, despite these differences, the fact remains that, as far as purely adiabatic dynamics are concerned, the Berry force is precisely the difference between semiclassical dynamics along $(i)$ along an eigenvalue of $\hat H$ and $(ii)$ along an eigenvalue of $\hat H^*$. Moreover, at the moment, in the nonadiabatic limit,  there is not a great deal of semiclassical understanding  about either $(i)$ how to treat Berry force effects or $(ii)$  how to predict differences between $\hat H$ and $\hat H^*$ dynamics in general.  For this reason, these two phenomena  are often discussed interchangeably -- as we will do here.  In the future, we hope it will be possible to more completely disentangle these phenomena (Berry force effects versus the difference between $\hat H$ and $\hat H^*$ dynamics) and develop a more precise, nuanced  semiclassical picture of  nuclear-electronic-spin dynamics. 

 As a side note, in all calculations that follow we use units such that $\hbar=1$.

\section{A Brief Review of Equilibrium and Nonequilibrium Fermi's Golden Rule (NEFGR)}

Often one is interested in the scenario whereby a vibronic system is initalized on  electronic state $\ket{1}$ and one interrogates the probability of reaching electronic state $\ket{2}$ at time $t$. Let $\hat\rho(0)$ be the initial density matrix, and we imagine partitioning the Hamiltonian as $\hat H = \hat H_0 + \hat V$, where $\hat H_0$ is the unperturbed Hamiltonian and $\hat V$ is the  perturbation; see Eq. \ref{eqveryfirst}.

According to standard practice, we transform to the interaction picture, perturbatively time-evolve the density matrix to first-order, and finally trace over the vibrational bath states $\{\textbf v'\}$ in the $\ket2$ electronic state. The resulting expression, which provides the 2-state population as a function of time, can be conveniently expressed using a time autocorrelation function $C(t',t'')$ of the interaction potential operator $\hat V_I(t) = e^{iH_0t}Ve^{-iH_0t}$:
\begin{gather}\label{FGR_time}
    P_2(t)\simeq\lvert V\rvert^2\int_0^t dt'\, \int_0^t dt''\, e^{-i(E_2-E_1)(t''-t')}C(t',t'')\\
    C(t',t'') = \text{Tr}_B\left[\hat\rho(0)\hat V_I(t'')\hat V_I(t')\right]\nonumber
\end{gather}
$\text{Tr}_B$ denotes a trace over the bath states.

At this juncture, according to a standard FGR calculation, one makes the approximation that the bath vibrational modes begin in thermal equilibrium. Thus, if the  vibrational states for state 1 are labeled $\{\ket{\textbf v}\}$, the initial density matrix is as follows:
\begin{equation}\label{rho0_eq}
    \hat\rho(0) = \sum_{\textbf v} e^{-\beta \hat H_B}\ket{1,\textbf v}\bra{1,\textbf v}
\end{equation}
Note that $\hat H_B$  acts only on the vibrational components of any eigenstate, and $\sum_\textbf v$ denotes a sum over all vibrational eigenstates of the bath. If one plugs Eq. \ref{rho0_eq} into Eq. \ref{FGR_time}, one finds that 
 correlation function reduces to a function dependent on only the difference $t'-t''$, and the final result in the energy domain is Eq. \ref{FGR}.  Note that evaluating either Eq. \ref{FGR} or Eq. \ref{FGR_time} 
 can be difficult, although the task is made much easier when we assume the bath is harmonic.

\begin{figure}[h]
    \centering
    \includegraphics[scale=0.5]{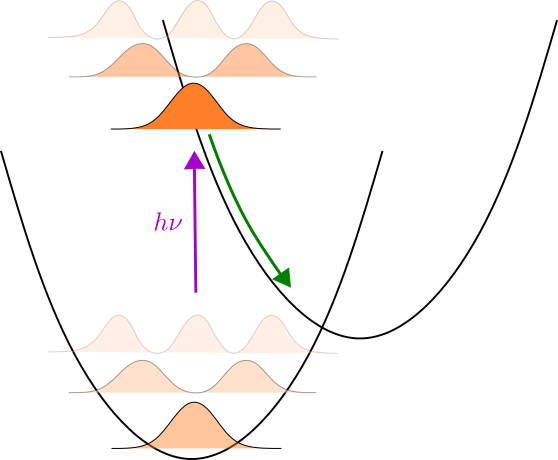}
    \caption{A fast photoexcitation is modeled as changing the electronic state without perturbing the bath coordinates. This transformation has the effect of placing the initial bath state out of equilibrium on the excited surface while preserving  the (canonical) vibrational density matrix in the Franck-Condon region.}  \label{fig_wp}
\end{figure}

Now, in order to generalize the FGR approach above to nonequilibrium initial conditions in a tractable manner, the usual prescription\cite{coalson:nitzan:1994:jcp:nonequilibrium, izmaylov:jcp:2011:nefgr} is to start with the equilibrium state above and then shift the eigenstates of each mode in space by a certain distance away from equilibrium. Mathematically, we can capture these initial conditions by applying to each bath eigenstate the following unitary translation operator:
\begin{equation}\label{translation}
    \hat{\textbf U}_{ne} = e^{-i\sum_\alpha d_\alpha\hat p_\alpha}
\end{equation}
This operation produces the nonequilibrium density matrix we will use as the initial state for subsequent calculations:
\begin{equation}\label{rho0_neq}
    \hat\rho_{neq}(0) = \sum_{\textbf v} e^{-i\sum_\alpha d_\alpha \hat p_\alpha}e^{-\beta \hat H_B}\ket{1,\textbf v}\bra{1,\textbf v}e^{i\sum_\alpha d_\alpha \hat p_\alpha}
\end{equation}
Eq. \ref{rho0_neq} represents a mixed state where each eigenstate has been shifted in space along each mode $\alpha$ by a distance $d_\alpha$. From here, following Refs. \cite{izmaylov:jcp:2011:nefgr}  
and 
\cite{coalson:nitzan:1994:jcp:nonequilibrium}, one can plug Eq. \ref{rho0_neq} into Eq. \ref{FGR_time} and use the same manipulations as in a typical FGR calculation to obtain a nonequilibrium expression.  Within this class of nonequilbrium initial conditions, one can model photophysical and photochemical experiments, for which it is reasonable to assume that the bath was originally thermalized in the ground electronic state before an electronic transition is driven. Thereafter, system dynamics are launched along an excited electronic state. See Fig. \ref{fig_wp}.

Finally, let us address the question of convergence. For perturbative results (based on equilibrium or nonequilibrium initial conditions) to be valid, it is generally sufficient that the dynamics induced by the perturbation correspond to the slowest time scale of the system being analyzed. For spin-boson and spin-boson-like problems, this condition requires that the nuclear dynamics not be significantly affected by the slow population leakage from one electronic state to the other, a condition which is dictated by the size of the perturbation and the frequencies of the bath. The validity of the NEFGR is {\em not} directly dictated by the size of the shift $d$ in Eq. \ref{rho0_neq}.\cite{nefgr_convergence}

\section{Theory}
Having reviewed the FGR/NEFGR formalisms, we will now apply the methods to the Hamiltonian with the coupling in Eq. \ref{ESOC}.
\subsection{Equilibrium dynamics}\label{Equilibrium_dynamics}
We begin in equililbrium. We will now demonstrate that the rate dynamics show no dependence on the sign of the terms $\left\{ W_\alpha \right\}$ if the initial state of the bath is a mixed state of vibrational eigenstates on either diabat. In other words, if the initial bath subspace density matrix is diagonal, there will be no Berry force effects in the population dynamics. Note that this statement is very strong and applies to many initial conditions; after all,  thermal equilibrium is just one special case of a diagonal bath density matrix.

To derive our result, we begin by applying a polaron transformation to the ESOC Hamiltonian, which is the typical approach in working with the standard spin-boson problem. The unitary operator which carries out the transformation is as follows:

\begin{equation}\label{polaron}
    \hat{\textbf U} = e^{-i\sum_\alpha\left(\lambda_\alpha^{(1)}\ket1\bra1+\lambda_\alpha^{(2)}\ket2\bra2\right)\hat p_\alpha}
\end{equation}
\begin{equation}
    \lambda_\alpha^{(j)} = \frac{c_\alpha^{(j)}}{m_\alpha\omega_\alpha^2}\nonumber
\end{equation}

In this new, so-called polaron representation, the diagonal electronic energies are decoupled from the boson bath; all coupling to the boson bath is captured purely in the interstate coupling. The exact form of the Hamiltonian is:
\begin{equation}\label{H_polaronic}
    \hat{\textbf U}\hat H\hat{\textbf U}^\dagger=\hat{\mathcal H} = \Tilde E_1\ket1\bra1+\Tilde E_2\ket2\bra2+\hat{\mathcal V}\ket1\bra2+\hat{\mathcal V}^\dagger\ket2\bra1+\hat H_B
\end{equation}
$$\hat{\mathcal V} = \mathcal Ve^{i\sum_\alpha  \lambda_\alpha \hat p_\alpha + W_\alpha \hat x_\alpha}\qquad \lambda_\alpha = \lambda_\alpha^{(2)}-\lambda_\alpha^{(1)}\qquad \Tilde E_j=E_j-\sum_\alpha\frac12m_\alpha\omega_\alpha^2(\lambda_\alpha^{(j)})^2$$
where $\mathcal V$ is a constant that differs from $V$ by a complex phase factor. The terms $\lambda_\alpha$ are the distances between the minima of states 1 and 2 along the alpha mode coordinates. Furthermore, we note that under the polaron transformation, eigenstates of either diabat are transformed to eigenstates of $\hat H_B$. In other words, the transformed initial density matrix is:
\begin{equation}
\label{eqrho_polaron}
\hat{\Tilde\rho}(0) = e^{-\beta\hat H_B} \ket{1}\bra{1}  
\end{equation}
In Eqs. \ref{H_polaronic} and \ref{eqrho_polaron}, we have added a superscript tilde over the energy $\tilde{E}_j$ and the density matrix $\hat{\tilde{\rho}}$, and used stylized $\mathcal{H}$ and $\mathcal{V}$,  so as to signify that these are quantities calculated after the initial polaron transformation.

Next we introduce the unitary operator
\begin{equation}\label{W_def}
    \hat{\mathcal R} = \prod_\alpha \text{exp}\left(i\left(\frac{\hat p^2_\alpha}{2m_\alpha}+\frac12m_\alpha\omega_\alpha^2\hat x_\alpha^2\right)\frac{\arctan\left(W_\alpha/\lambda_\alpha m_\alpha\omega_\alpha\right)}{\omega_\alpha}\right)
\end{equation}
which propagates each individual mode forward in time by an amount $$-(1/\omega_\alpha)\arctan(W_\alpha/\lambda_\alpha m_\alpha\omega_\alpha)$$
We note that, as far as each mode $(x_\alpha, p_{\alpha})$ is concerned,  $\hat{\mathcal R}$ is a function of the harmonic oscillator Hamiltonian, $\frac{\hat p_{\alpha}^2}{2m}+\frac12m\omega_{\alpha}^2\hat x_{\alpha}^2$.  Thus, $\hat{\mathcal R}$ preserves the eigenstates of $\hat H_B$, merely multiplying them by a phase factor.  Transforming $\hat{\mathcal H}$ under this operator leads to the (final) transformed Hamiltonian
\begin{equation}\label{W_trans}
    \hat{\mathscr H}=\hat{\mathcal R}\hat{\mathcal H}\hat{\mathcal R}^\dagger=\Tilde E_1\ket1\bra1+\Tilde E_2\ket2\bra2+e^{i\sum_\alpha\Gamma_\alpha \hat p_\alpha}\ket1\bra2+e^{-i\sum_\alpha\Gamma_\alpha \hat p_\alpha}\ket2\bra1+\hat H_B
\end{equation}
$$\Gamma_\alpha = \sqrt{\lambda_\alpha^2+\frac{W_\alpha^2}{m_\alpha^2\omega_\alpha^2}}$$
 At the same time, the initial density matrix is unchanged through this transformation:
\begin{equation}\label{rho0_W}
    \hat{\mathcal{R}}\hat{\Tilde\rho}(0)\hat{\mathcal{R}}^\dagger = \hat{\Tilde\rho}(0)=e^{-\beta \hat H_B} \ket{1}\bra{1}
\end{equation}
Further details of this  transformation (conjugation by $\hat{\mathcal R}$) are given in Appendix \ref{app_1}. 

In the end, the signs of the  $W_\alpha$ terms completely disappear when we transform both (i) the Hamiltonian and (ii) the equilibrium initial state.  Taking Eq. \ref{W_trans} and Eq. \ref{rho0_W} together, we find that (to any order) all vibronic dynamics depend only on the magnitudes of the terms $W_\alpha$, not on their signs: in fact, the form of $\hat{\mathscr H}$ suggests that the system has a dressed reorganization energy that can be expressed as a sum of dressed single-mode reorganization energies:
\begin{eqnarray}\label{lambda_sm}
    E_r^{(tot)} &=& \sum_{\alpha} E_{r,\alpha}^{(tot)} \\ 
    E_{r,\alpha}^{(tot)} & = &
    \frac12m_\alpha\omega_\alpha^2\Gamma_\alpha^2=\frac12m_\alpha\omega_\alpha^2\lambda_\alpha^2+\frac{W_\alpha^2}{2m_\alpha}
\end{eqnarray}

\begin{figure}[h]
    \centering
    \includegraphics[scale=0.5]{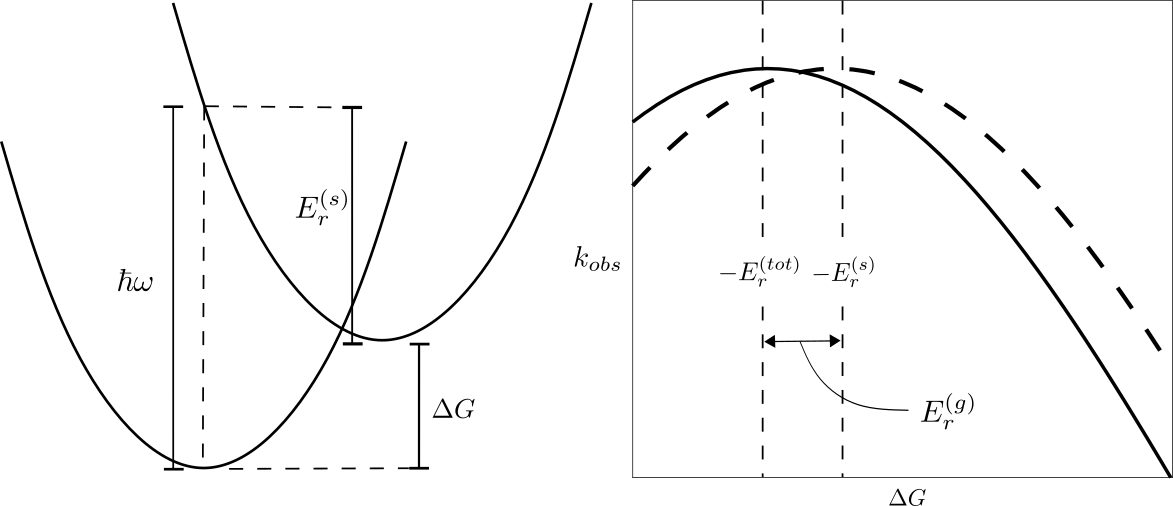}
    \caption{(Left) This schematic shows the relationship between the driving force $\Delta G$, the static reoganization energy $E_r^{(s)}$, and the vertical electronic excitation energy  $\hbar \omega$. (Right) Expected inverted regime electron transfer rate curves highlighting the difference between the static reorganization energy $E_r^{(s)}$ (dashed) and the total reorganization energy $E_r^{(tot)}$ (solid). For the Hamiltonian in Eq. \ref{ESOC}, the high temperature equilibrium electron transfer rate is given by Eq. \ref{hightemprate} with $s(t) = 0$ and where $E_r^{(tot)} > E_r^{(s)}$.}
    \label{E_r_fig}
\end{figure}

The dressed reorganization energy $E_{r}^{(tot)}$ is composed of two distinct components. The first is the standard reorganization energy, which we will henceforth refer to as $E_r^{(s)}$.
\begin{align}
    E_r^{(s)} &= \sum_{\alpha} \frac12m_\alpha\omega_\alpha^2\lambda_\alpha^2
\end{align}
The second is a term arising from geometric phase effects, which we will term $E_r^{(g)}$:
\begin{align}
    E_r^{(g)} &= \sum_{\alpha} \frac{W_\alpha^2}{2m_\alpha}
\end{align}
On the one hand, if one were to spectroscopically characterize the reorganization energy of a vibronic system with spin-orbit effects by measuring fluctuations in the energy gap, one would measure $E_r^{(s)}$. On the other hand, if one were to conduct a series of experiments for a particular electron-transfer system to construct a plot of $\Delta G$ vs. rate, one would observe a shift in the peak corresponding to $E_r^{(tot)}$. In other words, $E_r^{(g)}$ arises only when there is nuclear motion of interest.  With this physical interpretation in mind, we will term $E_r^{(tot)}$ the total  reorganization energy, $E_r^{(s)}$ the static reorganziation energy, and $E_r^{(g)}$ the geometric reorganization energy. See Fig. \ref{E_r_fig}.

Three key features of the Hamiltonian were necessary for the above proof. The first is that the bath modes were purely harmonic. The second is that the modes maintained their respective frequencies when there was a change in electronic state. Third and finally, we relied on the fact that the initial state was a mixed state of vibrational eigenstates of one of the two system diabats (which allowed us to write Eq. \ref{rho0_W}). As mentioned in the introduction, this state of affairs suggests  more distinct electron transfer rate effects may well emerge with either anharmonic bath models and/or nonequilibrium starting conditions. 

\subsection{Nonequilibrium dynamics}

In order to treat the nonequilibrium case using the formalism from above, the only change we must make is to modify the initial density matrix. We will use the density matrix defined in Eq. \ref{rho0_neq}. As shown in Appendix \ref{app_FGR}, the resulting autocorrelation function is:
\begin{align}
    C(t',t'')&=\text{Tr}\left[ e^{-\beta \hat H_B}e^{i\sum_\alpha d_\alpha \hat p_\alpha}e^{i\sum_\alpha \lambda_\alpha \hat p_\alpha(t'')+W_\alpha \hat x_\alpha(t'')}e^{-i\sum_\alpha\lambda_\alpha \hat p_\alpha(t')+W_\alpha \hat x_\alpha(t')}e^{-i\sum_\alpha d_\alpha \hat p_\alpha}\right]\\
    &=e^{-\sum_\alpha(\overline\lambda_\alpha^2+\overline W_\alpha^2)\left[2n_\alpha+1-(n_\alpha+1)e^{-i\omega_\alpha (t''-t')}-n_\alpha e^{i\omega_\alpha (t''-t')}\right]} \; \;   \times\nonumber\\
    &\qquad e^{\sum_\alpha\left(\overline\lambda_\alpha-i\overline W_\alpha\right)\left(e^{-i\omega_\alpha t''}-e^{-i\omega_\alpha t'}\right)\overline d_\alpha-\left(\overline\lambda_\alpha+i\overline W_\alpha\right)\left(e^{i\omega_\alpha t''}-e^{i\omega_\alpha t'}\right)\overline d_\alpha}\label{corr_func}\\
    &\qquad\overline{\lambda}_\alpha = \lambda_\alpha\sqrt{\frac{m_\alpha\omega_\alpha}2}\qquad \overline{W}_\alpha = \frac{W_\alpha}{\sqrt{2m_\alpha\omega_\alpha}}\qquad \overline{d}_\alpha = d_\alpha\sqrt{\frac{m_\alpha\omega_\alpha}2}\nonumber
\end{align}

Substituting this expression into the integral in Eq. \ref{FGR_time}, we obtain a general first-order population expression:
\begin{equation}\label{FGR_pop}
\begin{split}
     P_2(t)\simeq\lvert V\rvert^2\int_0^t dt'\, \int_0^t dt''\, e^{-i(E_2-E_1)(t''-t')}e^{-\sum_\alpha(\overline\lambda_\alpha^2+\overline W_\alpha^2)\left[2n_\alpha+1-(n_\alpha+1)e^{-i\omega_\alpha (t''-t')}-n_\alpha e^{i\omega_\alpha (t''-t')}\right]} \; \; \times \\
    \qquad e^{\sum_\alpha\left(\overline\lambda_\alpha-i\overline W_\alpha\right)\left(e^{-i\omega_\alpha t''}-e^{-i\omega_\alpha t'}\right)\overline d_\alpha-\left(\overline\lambda_\alpha+i\overline W_\alpha\right)\left(e^{i\omega_\alpha t''}-e^{i\omega_\alpha t'}\right)\overline d_\alpha}
\end{split}
\end{equation}

\begin{figure}
    \centering
    \includegraphics[scale = 0.5]{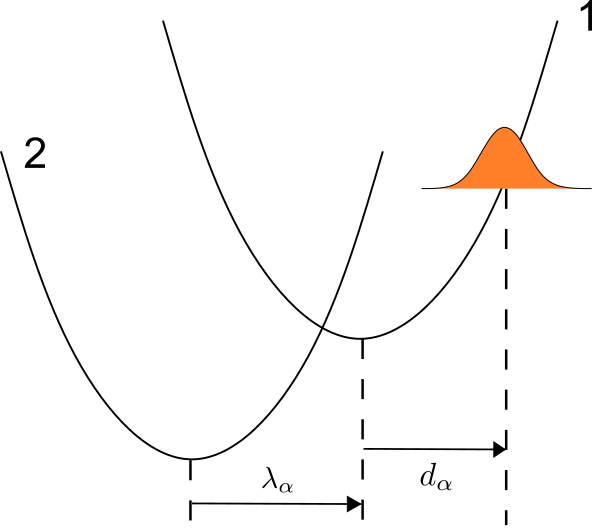}
    \caption{The physical meaning of the terms $\lambda_\alpha$ and $d_\alpha$.}
    \label{paramsfig}
\end{figure}

Here, $P_2(t)$ describes the gradual development of population in state $\ket 2$ starting with $P_2(0)=0$. One can take the time derivative of the equation above and obtain a time-dependent rate of transition from the $\ket 1$ state to the $\ket 2$ state.
The formal (time dependent) rate constant is given in Eq. \ref{formalrate}. In the high-temperature Markovian limit, the rate expression can be explicitly evaluated, producing a time-dependent golden rule rate expression of the form:
\begin{equation}\label{hightemprate}
    k(t)=\frac{2\pi\lvert V\rvert^2}{\sqrt{4\pi E_r^{(tot)} k_BT}}\text{exp}\left(-\frac{(\Delta G^\circ+E_r^{(tot)}+s(t))^2}{4E_r^{(tot)} k_BT}\right)
\end{equation}
\begin{equation}\label{dephasing_def}
    s(t) = \sum_\alpha d_\alpha\left[m_\alpha\omega_\alpha^2 \lambda_\alpha\cos(\omega_\alpha t)- W_\alpha\omega_\alpha\sin(\omega_\alpha t)\right]
\end{equation}
where $E_r^{(tot)}$ is the total reorganization energy that arises due to the combined effects of $E_r^{(s)}$ and $E_r^{(g)}$. Note that Eq. \ref{hightemprate} for the time-dependent rate constant has a form that is very similar to that of a classical Marcus theory time-independent rate expression. In this high temperature limit, the dressed reorganization energy $E_r^{(tot)}$ and dephasing function $s(t)$ are enough to fully characterize the nonequilibrium rate dynamics. These parameters are in turn characterized by the bath density of modes, as well as the shift parameters $\lambda_\alpha$, $ d_\alpha$, and $W_\alpha$. The physical meaning of each of these terms is summarized in Table \ref{constants} and Figure \ref{paramsfig} as a reminder to the reader.

\begin{table}
\begin{tabular}{|c|c|}
\hline
Parameter & Physical Meaning \\
\hline
$s(t)$ & Dephasing function for nonequilibrium fluctuations \\
\hline
$d_{\alpha}$ & Shift of initial bath states away from equilibrium \\
\hline
$W_\alpha$ & Spatial frequency of the phase oscillations of the interstate coupling \\
\hline
$\lambda_\alpha$ & Physical shift between diabatic potential wells \\
\hline
\end{tabular}
\caption{The key parameters that  define the rate of nonequilibrium relaxation.}\label{constants}
\end{table}

Let us now focus on the dephasing function $s(t)$ in Eq. \ref{dephasing_def}. This function contains all information regarding the transient behavior of the system, serving to capture the fluctuations in the interstate energy gap as the nuclear geometry evolves. Note in particular that the second term in $s(t)$ changes when any single $W_\alpha$ changes sign, clearly indicating a spin-dependent rate effect. Additionally, in the case that the number of modes in the bath is very large, $s(t)$ tends to 0 as $t$ grows large, implying that the system tends to equilibrium dynamics. Therefore, $s(t)$ indeed contains information about the dephasing of the bath. As the bath dephases and returns to equilibrium, we expect, based on the result from section \ref{Equilibrium_dynamics}, that all Berry force effects should  be damped out.

Before concluding this section and presenting our results, let us note that, in order to simulate photoexcitation in a two-state system, one can simply set
$ d_\alpha = -\lambda_\alpha$ for all $\alpha$ (see Fig. \ref{paramsfig}), and such a choice will manifest itself directly in the dynamics of the dephasing function in Eq. \ref{dephasing_def}.  Note, however, that 
the present model can also be applied to systems where the final acceptor state is not the same as the ground state, such as multistate models and/or intersystem crossings. In such cases, to retain full generality, one must not assume  $ d_\alpha=-\lambda_\alpha$.

\section{Results}

\subsection{Photoexcitation dynamics in a continuum bath model}

In a typical analysis of a spin-boson system (i.e. in the Condon approximation), one assumes that the system-bath interaction can be characterized by a spectral density. Working from such an approximation, one can compute the reorganization energy, which fully characterizes relaxation dynamics. However, in the model we have investigated thus far, a single spectral density is insufficient to characterize the system-bath interaction; the system exchanges energy through both the system-bath coupling and spin-orbit effects so that we must fix two sets of parameters, $\lambda_\alpha$ and $W_\alpha$.

To make such parameterization as simple as possible, we will assume a density of modes $\rho(\omega) = (\eta\omega/\omega_C^2)\exp(-\omega/\omega_C)$. $\omega_C$ describes a cutoff frequency for the bath and $\eta$ is a normalization factor that ensures integration over the density function returns the correct number of modes in the bath. For the diabats, we will then further assume that the mass-weighted shifts between the two surfaces are uniform across all modes, i.e. $\lambda_\alpha\sqrt{m_\alpha}=\lambda$ for all $\alpha$. Similarly, we set $W_\alpha/\sqrt{m_\alpha}$ equal to a constant $W$ for all $\alpha$. Note that $W$ can be positive or negative. For the bath displacement parameters $ d_\alpha$, we set the mass weighted displacements $ d_\alpha\sqrt{m_\alpha}$ equal to $-\kappa\lambda$. As discussed earlier, $\kappa=1$ corresponds to relaxation of a photoexcited system.

In order to characterize the dynamics, we need to compute the reorganization energy and the dephasing function. We begin with the reorganization energy. The total reorganization energy is the sum over all modes of the single-mode reorganization energies:
\begin{align}\label{photo_model_lambda}
    E_r^{(tot)} &= \sum_\alpha \frac12m_\alpha\omega_\alpha^2\lambda_\alpha^2+\frac{W_\alpha^2}{2m_\alpha}\nonumber\\
    & = \frac12\int_0^\infty d\omega\,\rho(\omega)(\omega^2\lambda^2+W^2)\nonumber \\
    & = 3\eta\omega_C^2\lambda^2+\frac\eta2W^2
\end{align}
Below, it will sometimes be helpful to switch variables from $(\lambda,W)$ to $(E_r^{(tot)}, \theta)$:
\begin{equation}\label{reparameter}
    \lambda = \sqrt{\frac{E_r^{(tot)}}{3\eta\omega_C^2}}\cos\theta \qquad W = \sqrt{\frac{2E_r^{(tot)}}{\eta}}\sin\theta
\end{equation}

Next, we calculate the dephasing function $s(t)$:
\begin{align}
    s(t) &=\sum_\alpha d_\alpha\left[m_\alpha\omega_\alpha^2 \lambda_\alpha\cos(\omega_\alpha t)- W_\alpha\omega_\alpha\sin(\omega_\alpha t)\right]\nonumber\\
    & = -\kappa\int_0^\infty d\omega\,\rho(\omega)\left(\omega^2\lambda^2\cos(\omega t)-\omega\lambda W\sin(\omega t)\right)\nonumber\\
    &=-\kappa\left[\frac{6\eta\lambda^2\omega_C^2\left(\omega_C^4t^4- 6\omega_C^2t^2+1\right)}{\left(\omega_C^2t^2+1\right)^4}+\frac{2\eta\lambda W\omega_C\left(\omega_C^3t^3-3\omega_Ct\right)}{\left(\omega_C^2t^2+1\right)^3}\right]\label{photo_model_s}
\end{align}
In terms of $E_r^{(tot)}$ and $\theta$, the dephasing function is:
\begin{equation}\label{photo_model_s_theta}
\begin{split}
        s(t) = -\frac{\kappa E_r^{(tot)}}{(\omega_C^2t^2+1)^4}\Big[(1+\cos2\theta)(\omega_C^4t^4-6\omega_C^2t^2+1)+\\
        \sqrt{\frac23}\sin2\theta(\omega_C^5t^5-2\omega_C^3t^3-3 \omega_Ct)\Big]
\end{split}
\end{equation}

\begin{figure}
    \centering
    \includegraphics[scale = 0.75]{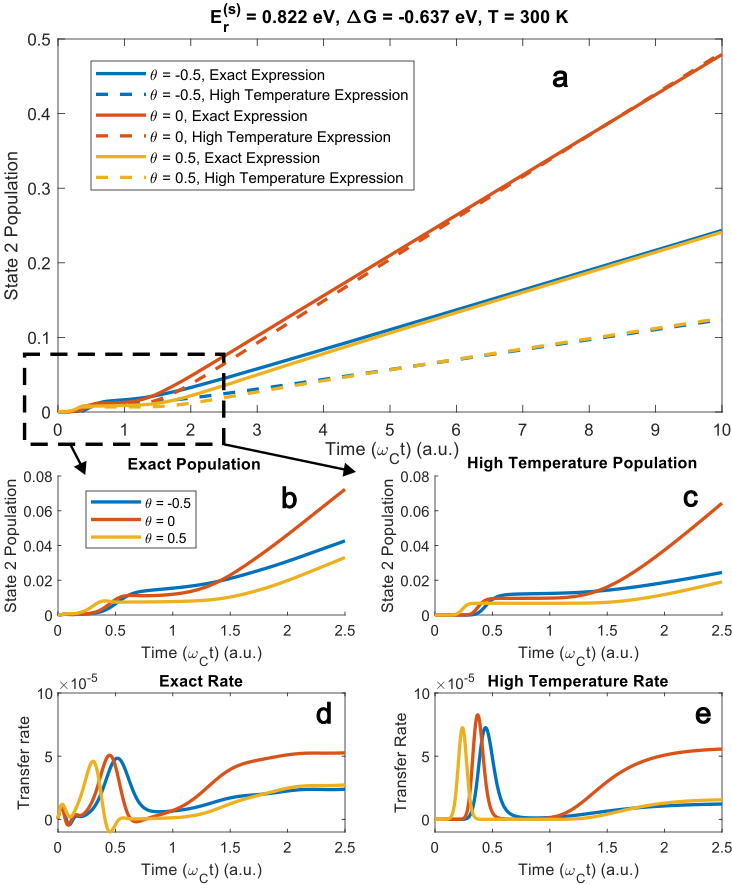}
    \caption{A comparison between the dynamics computed numerically using the exact Fermi Golden Rule expression and those derived from the high temperature expression for $\theta = 0, 0.5, -0.5$. We plot the population on state 2 $(P_2$) as a function of time.   $\omega_C$ is taken to be 0.001 a.u. For long times, the populations and rate expressions all converge to the equilibrium data, so that $P_2(\theta = 0.5) = P_2 (\theta = -0.5)$. At short times, however, as emphasized by subfigures $(b)-(e)$, the dynamics for $\pm  \theta$, i.e. $\pm W$, can be very difficult and exhibit strong Berry force effects.  Note that results computed using the high-temperature expression  qualitatively capture features of those computed with the exact FGR expression, especially at short times, tracking well the relative timing and magnitude of the transient rate spikes.}\label{hightemp_vs_exact}
\end{figure}

In Fig. \ref{hightemp_vs_exact}, we  show rate constants and population yields that were calculated for a photogenerated ($\kappa = 1$) distribution using the high-temperature expression,  Eqn. \ref{hightemprate}.
We fix the static reorganization energy $E_r^{(s)}$ and add in some amount of geometric reorganization energy ($E_r^{(g)} = \frac{W^2}{2m}$); the ratio is quantified by $\theta$ in Eq. \ref{reparameter}.
In subplot (a), we plot the population for short and long times for different values of $\theta$.
At long times, as expected, we see the dynamics for $\pm \theta$ gradually approach the same equilibrium rate as time progresses. Here, we emphasize that the disagreement between the exact NEFGR expression and the high temperature expression for $\theta = \pm 0.5$ is not the result of any interesting dynamical effect that depends on $\theta$.  Instead, a simple numerical calculation shows that the exact and high temperature expressions differ simply because of tunneling effects (whose importance depends on the total value of the reorganization energy $E_r^{(tot)}$). Note that $E_r^{(tot)(\theta = 0.5)} = 
E_r^{(tot)(\theta = -0.5)} \ne
E_r^{(tot)(\theta = 0)}$

The more interesting behavior occurs at short times while the bath still has yet to decohere and where the population curves can depend sensitively on $\pm \theta$ (i.e. $\pm W$). To see these dynamics clearly, we zoom in within subfigures $(b)$ and $(c)$. Here, we find transient populations for $\theta = -0.5$ that are 50\% larger than those of $\theta = 0.5$. In general, Berry force effects are larger when we calculate populations with the exact (rather than the high temperature) expression, but overall, the high temperature expression clearly reflects the correct qualitative physics (especially if we compare only $\theta$ with $-\theta$ data).  

Lastly, in subfigures $(d)$ and $(e)$, we plot the derivative of the population data, i.e. the instantaneous rates.  Here, we can clearly see differences in the features of the early time dynamics, including the magnitude and the relative timing of the populations spikes. Again, we find that the high temperature approximation is a reasonably good approximation for the exact early time dynamics (and computationally cheaper to evaluate). 

Finally, and most importantly, we would like to ask whether any rate differences between systems with opposite signs  ($ \pm W$) can persist for a significant period of time. 
In order to probe this question for a wide variety of $\theta$ values (while avoiding excessive computational expense), we will use the high-temperature expression above. We can roughly quantify dynamical differences  by focusing on rate dynamics at the time $1/\omega_C$, i.e.  the characteristic dephasing time of the bath.  For this data set, we will fix the total reorganization energy $E_r^{(tot)}$; if we were to fix $E_r^{(s)}$ as before, the variation in $E_r^{(tot)}$ would be very large and make it difficult to isolate effects arising from the geometric component. (Moreover, as noted above in Fig. \ref{hightemp_vs_exact}, the quantitative accuracy of the high-temperature approximation depends on the total value of $E_r^{(tot)}$ and whether one can safely ignore tunneling; by fixing $E_r^{(tot)}$, we can thus reasonably assume a cancellation of error and compare the effect of exchanging $E_r^{(g)}$ with $E_r^{(s)}$).

\begin{figure}[p]
    \centering
    \includegraphics[scale=0.6]{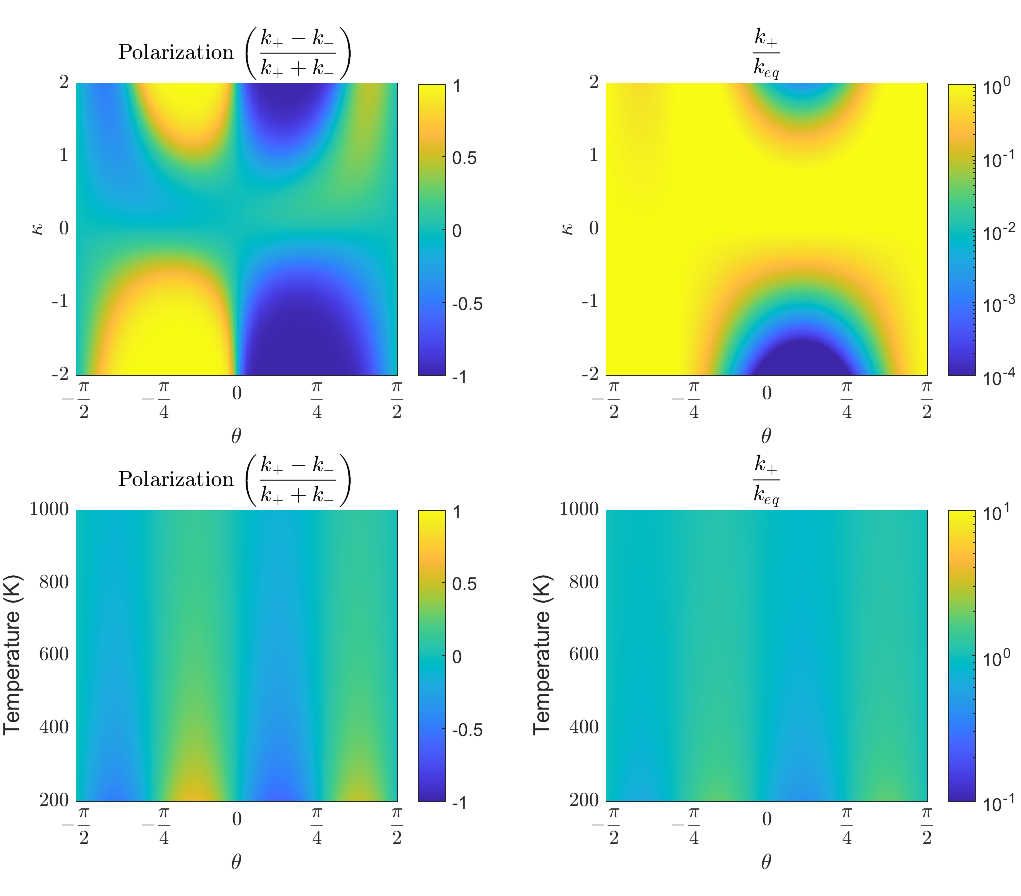}
    \caption{(Left)  Comparison of the spin polarization (see Eq. \ref{pm_ratio}) at time $1/\omega_C$ after initiation of dynamics for a range of values of $\kappa$, $\theta$, and temperature. Depending on the fraction of geometric vs. standard reorganization energy, one can find a great deal of spin polarization if one starts far from equilibrium (large $\kappa$) and if the temperature is not too large.  (Right) Comparison of the rate dynamics at time $1/\omega_C$ relative to the final equilibrium rate (following bath dephasing) for a range of values of $\kappa$, $\theta$, and temperature. Berry forces can have large effects, with relative transient rates $\sim 10^3$ possible between different $\left\{W_{\alpha}\right\}$ and $\left\{-W_{\alpha}\right\}$ values (leading to spin polarization on the order of unity). Parameters: All plots have $E_r^{(tot)} = 0.822$ eV and $\Delta G = -0.637$ eV. The upper plots were generated using a temperature of 300 K and the lower plots were generated for photoexcitations, i.e. $\kappa = 1$. All data was generated in the high temperature approximation (see Eqs. \ref{hightemprate} and \ref{photo_model_s_theta}).  }\label{rate_compare}
\end{figure}

Our results are plotted in Fig. \ref{rate_compare}. We consider two related quantities. First, we plot the spin polarization at time $1/\omega_C$:
\begin{equation}\label{pm_ratio}
    \frac{k_+-k_-}{k_++k_-}
\end{equation}
where $k(t)$ is as defined in Eq. \ref{hightemprate} and $k_-$ is related to $k_+$ through a change in sign on the $\{W_\alpha\}$ terms.
As shown in Fig. \ref{rate_compare}, the polarization can become quite large, reaching nearly 60 percent in photoexcitation processes in the 200 to 400 K range. Note the antisymmetry in the polarization plot across the line $\theta=0$. This antisymmetry arises naturally because changing the sign of $\theta$ corresponds to changing the signs on the $\{W_\alpha\}$ terms (which results in opposite spin polarization).

Second, in Fig. \ref{rate_compare}, we also plot
\begin{equation}\label{eq_ratio}
    \frac{k_+}{k_{eq}}
\end{equation}
in order to quantify the absolute magnitude of the effect of the $\left\{ W_{\alpha} \right\}$ terms.
We find that the nonequilibrium rates with $W \ne 0$ can be very different from the rates with $W=0$, sometimes by as much as a factor of 1000 for systems prepared far from equilibrium. Interestingly, the rate at time $1/\omega_C$ is never significantly faster than the equilibrium rate; but there are certain regimes in which the transfer rate is much slower. This result suggests that the spin polarization observed is a result not of favoring one electronic spin, but rather disfavoring the other. Finally, we observe that as the temperature of the system is raised, polarization effects due to spin-orbit effects gradually disappear, reflecting the fact that geometric phase effects are highly dependent on bath coherences. 

\begin{figure}[h]
    \centering
    \includegraphics[scale = 0.7]{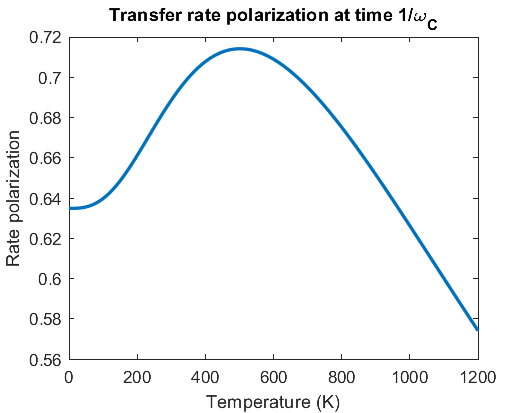}
    \caption{A visualization of the temperature dependence of the transient rate polarization (polarization at time $1/\omega_C$)  for $\theta=-0.5$. Note that the polarization attains a maximum at an intermediate temperature and approaches a fixed limit as the temperature approaches 0. (Although the high-temperature expression correctly captures the asymptotic behavior, the decrease in polarization at lower temperatures can only be observed using the full NEFGR expression.) Parameters: $E_r^{(s)}$ = 0.822 eV, $\Delta G$ = -0.637 eV, $\theta$ = -0.5.}
    \label{pol_vs_temp}
\end{figure}

Although the high temperature expression is useful, it cannot provide a complete picture of the system behavior at low to intermediate temperatures. In Fig. \ref{pol_vs_temp}, we plot the transient spin polarization as a function of temperature for $\theta = -0.5$ 
using the full NEFGR expression (not the high temperature expression).  We find a more nuanced picture; the spin polarization does not increase monotonically as one lowers the temperature, but rather has a maximum at some intermediate temperature. Interestingly, such non-monotonic temperature dependence has also been observed (experimentally and theoretically) by Naaman, Fransson and co-workers in the context of spin polarization arising from conduction through DNA base pairs \cite{Naaman_Fransson_CISS_2021_preprint}.  

\section{Discussion}

From the results presented above, it is clear that Berry forces can survive nuclear friction and that spin-dependent electron transfer dynamics are indeed possible. Although spin-orbit coupling effects (i.e. the effects of the sign of $W$) vanish in equilibrium and contribute only a small amount to the total reorganization energy, these effects can be quite important for systems prepared far from equilibrium. At the same time, however, the present work only scratches the surface of how electronic spin ties into electron transfer, and opens up a slew of questions.

First, what is the most reasonable Hamiltonian that one should use when describing spin-dependent electron transfer? In the present manuscript, we have simply assumed an exponential complex-valued diabatic coupling that changes according to $e^{i\sum_{\alpha}W_\alpha \hat x_\alpha}$.  Where does such a term come from? In principle, we expect an interstate coupling in the electronic Hamiltonian to be of the form:
    \begin{equation}\label{coupling}
        a(\hat x)+ib(\hat x) = c(\hat x)e^{id(\hat x)}
    \end{equation}
where $a(\hat x)$ is a standard, real-valued diabatic coupling and $ib(\hat x)$ is the purely imaginary-valued spin -orbit  coupling, as calculated in Appendix \ref{LS_app}.  Thus, the phase $d(\hat x)$ must originate from a competition between different effects (e.g. the ratio between a position dependent spin-orbit coupling and a position independent diabatic coupling, or the ratio between a position dependent diabatic coupling and a position independent spin-orbit coupling). If one assumes $c(\hat x)$ is a constant, and one expands  $d(\hat x) \approx \sum W_{\alpha} \hat x_{\alpha}$ as a Taylor series, one recovers the ESOC Hamiltonian in Eq. \ref{ESOC}.
At present, our group is expending a  great deal of effort running {\em ab initio} calculations searching for systems where  $|W|$ will be large and/or meaningful. One must wonder: how large can $|W|$ be in practice (especially if SOC is small), and is it reasonable to make the exponentiation in Eq. \ref{coupling} over the relevant volume of configuration space where two diabatic curves cross?

Questions about the validity of the ESOC model can be addressed in the future by investigating alternative forms for the interstate coupling, for instance the more common Linear Vibronic Coupling (LVC) type model\cite{izmaylov:jcp:2011:nefgr,koppel:review:conicalbook,yarkony:review:conicalbook}. In the LVC model, the interstate coupling is taken to be linear in position for all bath modes, i.e. $\hat V = \sum_\alpha c_\alpha\hat x_\alpha \ket{1} \bra{2} + c^*_\alpha\hat x_\alpha \ket{2} \bra{1}$. The LVC model is standard for investigating dynamics of systems around conical intersections, but this model must also break down at some point because the off-diagonal couplings grow to infinity away from the origin. Nevertheless, if one works with an LVC (or an LVC-like model), a second question one can ask is: what dynamics do we observe when a conical intersection is broken by a small amount of spin-orbit coupling? Note that, within an LVC model, adding spin-orbit coupling will be completely different from adding a real-valued constant coupling (of equal magnitude); a real valued coupling would shift the LVC conical intersection, whereas an imaginary spin-orbit coupling would remove the conical intersection entirely (and lead to a large $W$). While such a removal might also not seem entirely realistic\cite{truhlar:2003:likelihood_conical, truhlar:2011:comment_arasaki, truhlar:2018:transition_states}, this  scenario makes clear that electron transfer effects may emerge that are wholly unique to strongly spin-dependent systems.

A third question relates to the displaced harmonic oscillator model for nuclear motion.  Above, the  equivalence of equilibrium dynamics for $\hat H$ and $\hat H^*$
arose almost accidentally and is clearly tied to the ESOC Hamiltonian.  Although first-order FGR rates may be identical  for $\hat H$ and $\hat H^*$ for all rate processes (see Eq. \ref{FGR}), there is no reason to expect  (in general) that these equilibrium rates will be the same at second or higher order. For example, in Fig. \ref{morse} below, we plot  dynamics  for one single anharmonic mode, where dynamics are initialized in the lowest eigenstate of the upper diabat. Clearly, the dynamics show a small (but nonzero) spin dependence. In general, will these differences be small (because they show up only at higher orders of perturbation theory) or will these differences be significant (because they can add together for many anharmonic modes)? One must also be very curious about the dynamics of a model with Duschinskii rotations\cite{pollak:2004:duschinskii_cooling} -- do such rotations (which introduce stronger directionality on the bath reorganization) amplify the effects of spin-dependent nuclear forces?  One significant challenge will be to quantify the magnitude of spin-dependence for equilibrium dynamics in systems beyond the displaced oscillator approximation.

\begin{figure}[h]
    \centering
    \includegraphics[scale = 0.68]{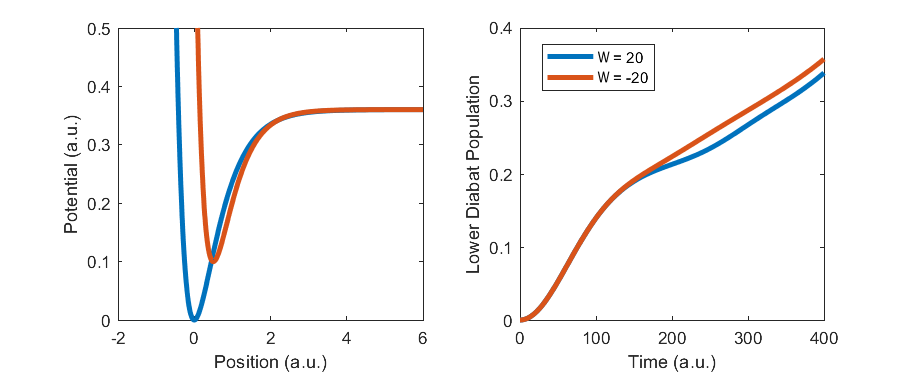}
    \caption{The Morse potential surfaces are parametrized such that they exhibit the same force constant in the harmonic approximation and the coupling between the surfaces is taken to be of the form $Ve^{iW\hat x} \ket{1} \bra{2} + V^*e^{-iW\hat x} \ket{2} \bra{1}$. The dynamics are initialized in the lowest eigenstate of the upper well. Note that the population dynamics for opposite signs on $W$ exhibit good agreement at very short time, but soon diverge, suggesting that higher order effects may be of great significance in anharmonic systems.}
    \label{morse}
\end{figure}

Finally, the goal of this research direction is to identify realistic systems that will display strong Berry force effects. 
Obviously, one would like to work with {\em ab initio} electronic structure Hamiltonians and run dynamics, but such a course of study is expensive. Even if, for a moment, we commit to studying model Hamiltonians, it is not wholly clear how to best map model dynamics to the real world. For instance, even though the standard spin-boson model has been used very successfully to model electron transfer over the years, it is clear that a more complicated parameter space of models will be necessary to discern Berry force effects.  After all, within the ansatz of a two level system, Berry force arises by breaking the Condon-approximation, and so we must require  not one, but two spectral densities; one of these spectral densities mediates energy exchange between the electronic and vibrational systems (as in the standard spin-boson model) and the other characterizes how the vibrational states mediate the spin-orbit coupling between electronic states. Do these interactions proceed with the same natural set of vibrational states? How should these spectra be tuned to one another in order to maximize spin-related electron transfer effects?  Can we disentangle these effects in a meaningful way? Lastly, if we return to the question of electronic structure, one must ponder how best to map a realistic system to the model Hamiltonians discussed above: in particular, will we necessarily find that, within floppy molecular structures, metal ions with large spin-orbit couplings always experience bigger Berry force effects during an electronic transition, or are the dynamics more complicated? For instance, some recent experiments suggest that the CISS effect can be sensitive to an ion with strong spin-orbit coupling\cite{torres-cavanillas:2020:spin_filtering} while other experiments suggest that spin-orbit coupling in a substrate is not important to the overall CISS signal\cite{zacharias:2018:jpcl:helicene_monolayers}.  If we can broadly answer such questions, we  will clearly gain a great deal of intuition as far as  rationally designing viable molecules and pathways with exciting new chemistry, and ideally gain the capacity to separate metal ions with large spin-orbit coupling from those with small spin-orbit coupling.

\section{Conclusions}

In conclusion, we have investigated a complex-valued spin-boson-like model Hamiltonian using nonequilibrium Fermi golden rule theory in order to provide fundamental insight into how spin-orbit coupling might influence condensed phase electron transfer. Our investigations have shown that Berry force effects can survive nuclear friction for some (measurable) period of time, before the effect dissipates.  The exact details of how long such an effect survives is complicated and depends sensitively on the exact nuclear motion -- the ESOC model presented here is simple to treat theoretically, but difficult to map to an {\em ab intio} Hamiltonian.  Further research  will  be necessary to better quantify the effect
and gain intuition as to exactly how big these Berry forces can be (and when) and if/how nuclear vibrations either correlate with or cause a CISS effect\cite{naaman:2012:jpcl}. Moreover, in the future, if we wish to run predictive simulations of realistic molecules with {\em ab initio} potentials, it will be necessary to develop efficient and inexpensive semiclassical algorithms\cite{wu:2020:neartotal,gonzalez:2015:ijqc_review,hoffmann_schatz:2000:jcp:isc}
for treating nuclear curve-crossing dynamics in the presence of spin orbit coupling.

\section{Appendix}
In this appendix, we briefly discuss the connection between electronic spin and complex Hamiltonians, and then proceed to derive some of the necessary equations presented above for the nonequilibrium golden rule calculations.

\subsection{Spin-dependence due to the $\hat{\textbf L}\cdot\hat{\textbf S}$ operator}\label{LS_app}

Consider a molecular system with two electronic states labeled $\ket1$ and $\ket2$ at a fixed geometry. In adding a spin-orbit effect, we must consider spin states $\ket{1,\uparrow}$, $\ket{2,\uparrow}$, $\ket{1,\downarrow}$, $\ket{2,\downarrow}$. In principle, we can write the Hamiltonian as a sum
\begin{equation}\label{hh}
    \hat H = \hat H_0+\xi\,\hat{\textbf{L}}\cdot\hat{\textbf S}
\end{equation}
where $\hat H_0$ is spin-independent and $\xi$ is the strength of the spin-orbit interaction. Since $\hat H_0$ is spin-independent, it can only couple states of identical spin. Therefore, in the spin-state basis, we can write $H_0$ as follows:
\begin{equation}
    \hat H_0 = \begin{pmatrix}
    E_1 & 0 & V & 0\\
    0 & E_1& 0 & V\\
    V & 0 & E_2 & 0\\
    0 & V & 0 & E_2
    \end{pmatrix}\qquad\begin{pmatrix}
    \ket{1,\uparrow}\\
    \ket{1,\downarrow}\\
    \ket{2,\uparrow}\\
    \ket{2,\downarrow}
    \end{pmatrix}
\end{equation}

Now we construct the $\hat{\textbf L}\cdot\hat{\textbf S}$ matrix. We note that if $\ket 1$ and $\ket 2$ are molecular orbitals in an appropriate representation, they do not couple to themselves through any of the three component angular momentum operators. Thus, we can write the $\hat{\textbf L}\cdot\hat{\textbf S}$ matrix as follows:
\begin{equation}
    \hat{\textbf L}\cdot\hat{\textbf S} = \begin{pmatrix}
    0 & 0 & \Braket{1}{\hat{\textbf L}_z}{2} & \Braket{1}{\hat{\textbf L}_x-i\hat{\textbf L}_y}{2}\\
    0 & 0 & \Braket{1}{\hat{\textbf L}_x+i\hat{\textbf L}_y}{2} & -\Braket{1}{\hat{\textbf L}_z}{2}\\
    \Braket{2}{\hat{\textbf L}_z}{1} & \Braket{2}{\hat{\textbf L}_x-i\hat{\textbf L}_y}{1} & 0 & 0\\
    \Braket{2}{\hat{\textbf L}_x+i\hat{\textbf L}_y}{1} & -\Braket{1}{\hat{\textbf L}_z}{2} & 0 & 0
    \end{pmatrix}
\end{equation}
Noting that the angular momentum component operators are purely imaginary, we can write the matrix more concisely:
\begin{equation}
    \hat{\textbf L}\cdot\hat{\textbf S} = \begin{pmatrix}
    0 & 0 & \alpha & \beta\\
    0 & 0 & -\beta^*& -\alpha\\
    \alpha^* & -\beta & 0 & 0\\
    \beta^* & -\alpha^* & 0 & 0
    \end{pmatrix}\qquad\begin{pmatrix}
    \ket{1,\uparrow}\\
    \ket{1,\downarrow}\\
    \ket{2,\uparrow}\\
    \ket{2,\downarrow}
    \end{pmatrix}
\end{equation}
Note that $\alpha$ is purely imaginary whereas $\beta$ has a real component, so the upper right block is anti-Hermitian and traceless. Therefore, this 2x2 matrix is diagonalizable and has purely imaginary eigenvalues $\alpha'$ and $\alpha'^*=-\alpha'$. The lower left block is just the conjugate transpose of the (anti-Hermitian) upper right block, and so is diagonalized by the same change of basis. Furthermore, the eigenvalues of the lower left block are conjugate to the eigenvalues of the upper right block. Note that a change of spin basis does not affect $\hat H_0$, which is block diagonal. We can now write down the total fixed-geometry Hamiltonian in the new spin basis (recall that $\xi$ is defined in Eq. \ref{hh}):
\begin{equation}
    \hat H = \begin{pmatrix}
    E_1 & 0 & V & 0\\
    0 & E_1 & 0 & V\\
    V &  0 & E_2 & 0\\
    0 & V & 0 & E_2
    \end{pmatrix}+\begin{pmatrix}
    0 & 0 & \xi\cdot\alpha' & 0\\
    0 & 0& 0 & \xi\cdot\alpha'^*\\
    \xi\cdot\alpha'^* & 0 & 0 & 0\\
    0 & \xi\cdot\alpha' & 0 & 0
    \end{pmatrix}\qquad\begin{pmatrix}
    \ket{1,\uparrow'}\\
    \ket{1,\downarrow'}\\
    \ket{2,\uparrow'}\\
    \ket{2,\downarrow'}
    \end{pmatrix}
\end{equation}
Therefore, at any nuclear geometry, the four-dimensional
Hilbert space above (for spin-electronic wavefunctions with two spatial orbitals) can always be partitioned into two distinct subspaces, each with a distinct Hamiltonian $\hat H$ and $\hat H^*$ corresponding to one spin or the other opposite spin.  

In this article, we have made the assumption that that such a partition will be valid and unchanged for all nuclear geometries.  This strong assumption allows us to assume that we can entirely ignore all spin-flip processes.

\subsection{Evaluation of Eq. \ref{W_trans}}\label{app_1}
To evaluate Eq. \ref{W_trans}, we begin with the polaronic Hamiltonian in \ref{H_polaronic}, restated here:
\begin{equation*}
    \hat{H} = E_1\ket1\bra1+ E_2\ket2\bra2+\hat{V}\ket1\bra2+\hat{V}^\dagger\ket2\bra1+\hat H_B
\end{equation*}
$$\hat{V} = Ve^{i\sum_\alpha  \lambda_\alpha \hat p_\alpha + W_\alpha \hat x_\alpha}\qquad \lambda_\alpha = \frac{c_\alpha^{(1)}-c_\alpha^{(2)}}{m_\alpha\omega_\alpha^2}$$

We will use $\hat{\mathcal R}$ as defined in Eq. \ref{W_def} to perform a unitary transformation on the polaronic Hamiltonian. We need to compute $\hat{\mathcal R}\hat H\hat{\mathcal R}^\dagger$. $\hat{\mathcal R}$ obviously commutes with $\hat H_B$. Thus, all we need to compute is $\hat{\mathcal R}e^{i\sum_\alpha W_\alpha \hat x_\alpha + \lambda_\alpha \hat p_\alpha}\hat{\mathcal R}^\dagger$. Using the fact that (for any operator $\hat M$)
\begin{align}
    e^{i\hat Ht} e^{\hat M} e^{-i\hat Ht} = e^{M(t)}
\end{align}
it follows that:
\begin{align}
    \hat{\mathcal R} e^{i\sum_\alpha W_\alpha \hat x_\alpha+\lambda_\alpha \hat p_\alpha}\hat{\mathcal R}^\dagger=\prod_\alpha e^{i\left( W_\alpha \hat x_\alpha(t)+\lambda_\alpha \hat p_\alpha(t)\right)}\Big\vert_{t=\arctan\left(\overline W_\alpha/\overline\lambda_\alpha\right)/\omega_\alpha}
\end{align}
\begin{equation*}
    \overline{\lambda}_\alpha = \lambda_\alpha\sqrt{\frac{m_\alpha\omega_\alpha}2}\qquad \overline{W}_\alpha = \frac{W_\alpha}{\sqrt{2m_\alpha\omega_\alpha}}
\end{equation*}

Keeping in mind that $t=\arctan\left(\overline W_\alpha/\overline \lambda_\alpha\right)/\omega_\alpha$, we find:
\begin{align}
    i\left( W_\alpha \hat x_\alpha(t)+\lambda_\alpha \hat p_\alpha(t)\right)&=\overline\lambda_\alpha\left(ae^{-i\omega_\alpha t}-a^\dagger e^{i\omega_\alpha t}\right)+i\overline W_\alpha\left(ae^{-i\omega_\alpha t}+a^\dagger e^{i\omega_\alpha t}\right)\\
    &=\frac{\left(\overline\lambda_\alpha+i\overline W_\alpha\right)\left(\overline\lambda_\alpha-i\overline W_\alpha\right)}{\sqrt{\overline\lambda_\alpha^2+\overline W_\alpha^2}}a-\frac{\left(\overline\lambda_\alpha-i\overline W_\alpha\right)\left(\overline\lambda_\alpha+i\overline W_\alpha\right)}{\sqrt{\overline\lambda_\alpha^2+\overline W_\alpha^2}}a^\dagger\\
    &=i\left(\sqrt{\frac{2\left(\overline\lambda_\alpha^2+\overline W_\alpha^2\right)}{m_\alpha\omega_\alpha}}\right)\hat p_\alpha = i\left(\sqrt{\lambda_\alpha^2+\frac{W_\alpha^2}{m_\alpha^2\omega_\alpha^2}}\right)\hat p_\alpha=i\Gamma_\alpha \hat p_\alpha
\end{align}

Therefore, we conclude that 
\begin{equation}
    \hat{\mathcal R}\hat H\hat{\mathcal R}^\dagger = E_1\ket1\bra1+E_2\ket2\bra2+Ve^{i\sum_\alpha\Gamma_\alpha \hat p_\alpha}\ket1\bra2+V^*e^{-i\sum_\alpha\Gamma_\alpha \hat p_\alpha}\ket2\bra1+\hat H_B
\end{equation}

\subsection{Derivation of Eq. \ref{FGR_pop}}\label{app_FGR}
In order to derive Eq. \ref{FGR_pop}, we will work in the polaron representation of the ESOC Hamiltonian, as described above. We write the Hamiltonian as
\begin{gather}
    \hat{H} = \hat H_0 + \hat V\\
    \hat H_0 = E_1\ket1\bra1+E_2\ket2\bra2+\hat H_B\\
    \hat V = Ve^{i\sum_\alpha  \lambda_\alpha \hat p_\alpha + W_\alpha \hat x_\alpha}\ket1\bra2+V^*e^{-i\sum_\alpha  \lambda_\alpha \hat p_\alpha + W_\alpha \hat x_\alpha}\ket2\bra1
\end{gather}

Beginning with the density matrix $\hat \rho(0)$ in (2.4), We would like to approximate
\begin{align}
    \sum_{\textbf v'}\Braket{2,\textbf v'}{\rho(t)}{2,\textbf v'}&=\sum_{\textbf v'}\sum_{\textbf v}\Braket{2,\textbf v'}{e^{-i\hat Ht}e^{-i\sum_\alpha d_\alpha \hat p_\alpha}e^{-\beta \hat H_B}}{1,\textbf v}\Braket{1,\textbf v}{e^{i\sum_\alpha  d_\alpha \hat p_\alpha}e^{i\hat Ht}}{2,\textbf v'}\\
    &=\sum_{\textbf v}\Braket{1,\textbf v}{e^{i\sum_\alpha  d_\alpha \hat p_\alpha}e^{i\hat Ht}}{2}\Braket{2}{e^{-i\hat Ht}e^{-i\sum_\alpha d_\alpha \hat p_\alpha}e^{-\beta \hat H_B}}{1,\textbf v}\\
    &=\text{Tr}_B\left[\Braket{1}{e^{i\sum_\alpha  d_\alpha \hat p_\alpha}e^{i\hat Ht}}{2}\Braket{2}{e^{-i\hat Ht}e^{-i\sum_\alpha d_\alpha \hat p_\alpha}}{1}e^{-\beta \hat H_B}\right]\\
    &=\text{Tr}_B\left[e^{i\sum_\alpha d_\alpha \hat p_\alpha}\Braket{1}{e^{i\hat Ht}}{2}\Braket{2}{e^{-i\hat Ht}}{1}e^{-i\sum_\alpha d_\alpha \hat p_\alpha} e^{-\beta \hat H_B} \right]
    \label{intrho2}
\end{align}

Using the definitions $\hat H_0$ and $\hat V$ from above, we expand $\exp(-i\hat Ht)$ to first order in a Dyson series as follows (\textbf{I} is the identity operator):
\begin{align}
    \Braket*{2}{e^{-i\hat Ht}}{1}&\simeq \Braket*{2}{\textbf I -i\int_0^tdt'\,e^{i\hat H_0t'}\hat Ve^{-i\hat H_0t'}}{1}\nonumber\\
    &=-iV^*\int_0^tdt'\,e^{i(E_2-E_1)t'}e^{i\hat H_Bt'}e^{-i\sum_\alpha\lambda_\alpha \hat p_\alpha+W_\alpha \hat x_\alpha}e^{-i\hat H_Bt'}
    \label{U21}
\end{align}

Now, we substitute this expression for $\Braket*{2}{e^{-i\hat Ht}}{1}$ (as well as the Hermitian transpose  $\Braket*{1}{e^{i\hat Ht}}{2}$) into Eq. \ref{intrho2}. Recognizing that $\text{Tr}_B\left[\exp(-\beta\hat H_B)\hat A\right]=\left\langle\hat A\right\rangle$ for an operator $\hat A$, we find that (in the end), the population becomes a  double integral with a time correlation function that must be evaluated:
\begin{gather}
    \sum_{\textbf v'}\Braket{2,\textbf v'}{\rho(t)}{2,\textbf v'}\simeq\lvert V\rvert^2\int_0^t dt'\, \int_0^t dt''\, e^{-i(E_2-E_1)(t''-t')}C(t',t'')\\
    C(t',t'')=\Big\langle e^{i\sum_\alpha d_\alpha \hat p_\alpha}e^{i\sum_\alpha \lambda_\alpha \hat p_\alpha(t'')+W_\alpha \hat x_\alpha(t'')}e^{-i\sum_\alpha\lambda_\alpha \hat p_\alpha(t')+ W_\alpha \hat x_\alpha(t')}e^{-i\sum_\alpha  d_\alpha \hat p_\alpha}\Big\rangle
\end{gather}

In order to evaluate the time correlation function in the integrand, it is a relatively straightforward application of two operator identities:
\begin{equation*}
    e^{\hat A}e^{\hat B}=e^{\hat A+\hat B+\frac12\left[\hat A,\hat B\right]}\qquad\left\langle e^{\hat A}\right\rangle = e^{\frac12\left\langle\hat A^2\right\rangle}
\end{equation*}
These relations are valid for operators $\hat A$ and $\hat B$ that linear in the position and momentum operators, which is the case for the product operator in the expectation value above.

As before, we use the following reduced forms for the coupling and shift coordinates:
\begin{equation*}
    \overline{\lambda}_\alpha = \lambda_\alpha\sqrt{\frac{m_\alpha\omega_\alpha}2}\qquad\overline{d}_\alpha =  d_\alpha\sqrt{\frac{m_\alpha\omega_\alpha}2}\qquad \overline{W}_\alpha = \frac{W_\alpha}{\sqrt{2m_\alpha\omega_\alpha}}
\end{equation*}

We apply the first identity repeatedly to obtain
\begin{align}
    &C(t',t'') = e^{\sum_\alpha\left(\overline\lambda_\alpha-i\overline W_\alpha\right)\left(e^{-i\omega_\alpha t''}-e^{-i\omega_\alpha t'}\right)\overline d_\alpha-\left(\overline\lambda_\alpha+i\overline W_\alpha\right)\left(e^{i\omega_\alpha t''}-e^{i\omega_\alpha t'}\right)\overline d_\alpha}\times\nonumber\\
    &e^{-\frac12\sum_\alpha\left(\overline\lambda_\alpha^2+\overline W_\alpha^2\right)\left(e^{i\omega_\alpha(t''-t')}-e^{-i\omega_\alpha(t''-t')}\right)}\Big\langle e^{\sum_\alpha\left(\overline\lambda_\alpha+i\overline W_\alpha\right)\left(e^{i\omega_\alpha t''}-e^{i\omega_\alpha t'}\right)a_\alpha^\dagger-\left(\overline\lambda_\alpha-i\overline W_\alpha\right)\left(e^{-i\omega_\alpha t''}-e^{-i\omega_\alpha t'}\right)a_\alpha}\Big\rangle
\end{align}

Finally, we apply the second identity to obtain the desired result:
\begin{align}
    C(t',t'')=&e^{-\sum_\alpha(\overline\lambda_\alpha^2+\overline W_\alpha^2)\left[2n_\alpha+1-(n_\alpha+1)e^{-i\omega_\alpha (t''-t')}-n_\alpha e^{i\omega_\alpha (t''-t')}\right]}\times\nonumber\\
    &\qquad e^{\sum_\alpha\left(\overline\lambda_\alpha-i\overline W_\alpha\right)\left(e^{-i\omega_\alpha t''}-e^{-i\omega_\alpha t'}\right)\overline d_\alpha-\left(\overline\lambda_\alpha+i\overline W_\alpha\right)\left(e^{i\omega_\alpha t''}-e^{i\omega_\alpha t'}\right)\overline d_\alpha}
\end{align}

\subsection{The High-Temperature Rate: Derivation of Eq. \ref{hightemprate}}\label{app_hightemp}
To get a time-dependent rate expression, we take the time derivative of the double integral describing the 2-state population:
\begin{align}
    \pd{}{t}\lvert V\rvert^2\int_0^t dt'\, \int_0^t dt''\, e^{-i(E_2-E_1)(t''-t')}C(t',t'')=&\lvert V\rvert^2\int_0^t dt''\, e^{-i(E_2-E_1)(t-t')}C(t',t)+\nonumber\\
    &\lvert V\rvert^2\int_0^t dt'\, e^{-i(E_2-E_1)(t''-t)}C(t,t'')
    \label{formalrate}
\end{align}

We begin by focusing on the first integral in Eq. \ref{formalrate}:
\begin{align}
    \lvert V\rvert^2\int_0^t dt''\, e^{-i(E_2-E_1)(t''-t)}&e^{-\sum_\alpha(\overline\lambda_\alpha^2+\overline W_\alpha^2)\left[2n_\alpha+1-(n_\alpha+1)e^{-i\omega_\alpha (t''-t)}-n_\alpha e^{i\omega_\alpha (t''-t)}\right]}\times\nonumber\\
    & e^{\sum_\alpha\left(\overline\lambda_\alpha-i\overline W_\alpha\right)\left(e^{-i\omega_\alpha t''}-e^{-i\omega_\alpha t}\right)\overline d_\alpha-\left(\overline\lambda_\alpha+i\overline W_\alpha\right)\left(e^{i\omega_\alpha t''}-e^{i\omega_\alpha t}\right)\overline d_\alpha}\\
    =\lvert V\rvert^2\int_0^t dt''\, e^{-i(E_2-E_1)(t''-t)}&e^{-\sum_\alpha(\overline\lambda_\alpha^2+\overline W_\alpha^2)\left[2n_\alpha+1-(n_\alpha+1)e^{-i\omega_\alpha (t''-t)}-n_\alpha e^{i\omega_\alpha (t''-t)}\right]}\times\nonumber\\
    & e^{\sum_\alpha\left(\overline\lambda_\alpha-i\overline W_\alpha\right)e^{-i\omega_\alpha t}\left(e^{-i\omega_\alpha (t''-t)}-1\right)\overline d_\alpha-\left(\overline\lambda_\alpha+i\overline W_\alpha\right)e^{i\omega_\alpha t}\left(e^{i\omega_\alpha (t''-t)}-1\right)\overline d_\alpha}\\
    =\lvert V\rvert^2\int_{-t}^0 d\tau\, e^{-i(E_2-E_1)\tau}&e^{-\sum_\alpha(\overline\lambda_\alpha^2+\overline W_\alpha^2)\left[2n_\alpha+1-(n_\alpha+1)e^{-i\omega_\alpha \tau}-n_\alpha e^{i\omega_\alpha \tau}\right]}\times\nonumber\\
    & e^{\sum_\alpha\left(\overline\lambda_\alpha-i\overline W_\alpha\right)e^{-i\omega_\alpha t}\left(e^{-i\omega_\alpha \tau}-1\right)\overline d_\alpha-\left(\overline\lambda_\alpha+i\overline W_\alpha\right)e^{i\omega_\alpha t}\left(e^{i\omega_\alpha \tau}-1\right)\overline d_\alpha}
\end{align}

In the high temperature limit, the second term in the integral decays quickly and we can use small-$\tau$ approximations. Therefore, we can approximate the integral as
\begin{align}
    \lvert V\rvert^2\int_{-t}^0 d\tau\,e^{-i(E_2-E_1)\tau}&e^{-\sum_\alpha(\overline\lambda_\alpha^2+\overline W_\alpha^2)\left[2n_\alpha+1-(n_\alpha+1)e^{-i\omega_\alpha \tau}-n_\alpha e^{i\omega_\alpha \tau}\right]}\times\nonumber\\
    & e^{-\sum_\alpha\left(\overline\lambda_\alpha-i\overline W_\alpha\right)e^{-i\omega_\alpha t}\left(i\omega_\alpha\tau\right)\overline d_\alpha+\left(\overline\lambda_\alpha+i\overline W_\alpha\right)e^{i\omega_\alpha t}\left(i\omega_\alpha \tau\right)\overline d_\alpha}\\
    =\lvert V\rvert^2\int_{-t}^0 d\tau\,e^{-i(E_2-E_1)\tau}&e^{-\sum_\alpha(\overline\lambda_\alpha^2+\overline W_\alpha^2)\left[2n_\alpha+1-(n_\alpha+1)e^{-i\omega_\alpha \tau}-n_\alpha e^{i\omega_\alpha \tau}\right]}\times\nonumber\\
    & e^{-\sum_\alpha 2i\omega_\alpha\overline d_\alpha\left[\overline\lambda_\alpha\cos(\omega_\alpha t)-\overline W_\alpha\sin(\omega_\alpha t)\right]\tau}
\end{align}

Next, we address the second integral in Eq. \ref{formalrate}:
\begin{align}
    \lvert V\rvert^2\int_0^t d\tau\,e^{-i(E_2-E_1)\tau}&e^{-\sum_\alpha(\overline\lambda_\alpha^2+\overline W_\alpha^2)\left[2n_\alpha+1-(n_\alpha+1)e^{-i\omega_\alpha \tau}-n_\alpha e^{i\omega_\alpha \tau}\right]}\times\nonumber\\
    & e^{-\sum_\alpha 2i\omega_\alpha\overline d_\alpha\left[\overline\lambda_\alpha\cos(\omega_\alpha t)-\overline W_\alpha\sin(\omega_\alpha t)\right]\tau}
\end{align}

Putting it all together and letting the bounds of the integral tend to infinity (invoking the Markovian approximation), we obtain the time dependent rate expression
\begin{equation}
    k(t)=\lvert V\rvert^2\int_{-\infty}^\infty d\tau\,e^{-i(E_2-E_1+s(t))\tau}e^{-\sum_\alpha(\overline\lambda_\alpha^2+\overline W_\alpha^2)\left[2n_\alpha+1-(n_\alpha+1)e^{-i\omega_\alpha \tau}-n_\alpha e^{i\omega_\alpha \tau}\right]}
\end{equation}
where $s(t) = \sum_\alpha2\omega_\alpha\overline d_\alpha\left[\overline\lambda_\alpha\cos(\omega_\alpha t)-\overline W_\alpha\sin(\omega_\alpha t)\right]$. This is just the Franck-Condon rate expression with a time fluctuation on the energy gap; in the high-temperature limit, this integral evaluates to
\begin{align}
    k(t)&=\frac{2\pi\lvert V\rvert^2}{\sqrt{4\pi E_r^{(tot)} k_BT}}\text{exp}\left(-\frac{(E_1-E_2-s(t)-E_r^{(tot)})^2}{4E_r^{(tot)} k_BT}\right)\\
    &=\frac{2\pi\lvert V\rvert^2}{\sqrt{4\pi E_r^{(tot)} k_BT}}\text{exp}\left(-\frac{(\Delta G^\circ+s(t)+E_r^{(tot)})^2}{4E_r^{(tot)} k_BT}\right)
\end{align}
where $E_r^{(tot)}$ describes a total reorganization energy due to the combined effects of the system-bath coupling and interstate coupling phase.
\section*{Acknowledgements}
This work was funded by the Center for Sustainable Separation of Metals, an NSF Center for Chemical Innovation (CCI) Grant CHE-1925708.

\bibliography{finalbib}

\begin{mcitethebibliography}{54}
\bibitem[nef()]{nefgr_convergence}
From Eq. \ref{hightemprate}, the nonequilibrium shift of the bath enters only
  in the dephasing function $s(t)$. Thus, regardless of how far from
  equilibrium the bath is, the electron transfer rate is bounded by
  \begin{equation} \frac{2\pi\lvert V\rvert^2}{\sqrt{4\pi E_r^{(tot)}k_BT}}
  \end{equation} Therefore, for this particular model, whenever an equilibrium
  Fermi Golden Rule dynamics is valid for a thermalized bath, a nonequililbrium
  FGR calculation should also be valid -- even for large initial displacements
  from equilibrium.\relax
\mciteBstWouldAddEndPunctfalse
\mciteSetBstMidEndSepPunct{\mcitedefaultmidpunct}
{}{\mcitedefaultseppunct}\relax
\EndOfBibitem
\end{mcitethebibliography}
\end{document}